\documentclass[preprint]{elsarticle}

\usepackage{hyperref}
\hypersetup{colorlinks=true}

\journal{arXiv.org cs.HC}
\biboptions{authoryear}

\usepackage{tabu}                      
\usepackage{booktabs}                  
\usepackage{gensymb}                   

\usepackage{fancyhdr} 
\usepackage{lastpage} 
\usepackage{xpatch}
\xpatchcmd{\pprintMaketitle}{\hrule}{}{}{} 
\xpatchcmd{\pprintMaketitle}{\hrule}{}{}{} 

\usepackage{soul} 

\begin{document}

\begin{frontmatter}

\title{Designing a 3D Gestural Interface to Support User Interaction with Time-Oriented Data as Immersive 3D Radar Chart}

\author[vrxaraddress,ivisaddress]{Nico Reski\corref{correspondingauthor}}
\ead{nico.reski@liu.se}

\author[vrxaraddress]{Aris Alissandrakis\corref{correspondingauthor}}
\cortext[correspondingauthor]{Corresponding author}
\ead{aris.alissandrakis@lnu.se}

\author[isovisaddress,ivisaddress]{Andreas Kerren}
\ead{andreas.kerren@liu.se}

\address[vrxaraddress]{VRxAR Labs, Department of Computer Science and Media Technology, Linn{\ae}us University, V\"axj\"o, Sweden}
\address[isovisaddress]{ISOVIS, Department of Computer Science and Media Technology, Linn{\ae}us University, V\"axj\"o, Sweden}
\address[ivisaddress]{iVis, Department of Science and Technology, Link\"oping University, Norrk\"oping, Sweden}

\begin{abstract}
The design of intuitive three-dimensional user interfaces is vital for interaction in virtual reality, allowing to effectively close the loop between a human user and the virtual environment.
The utilization of 3D gestural input allows for useful hand interaction with virtual content by directly grasping visible objects, or through invisible gestural commands that are associated with corresponding features in the immersive 3D space.
The design of such interfaces remains complex and challenging.
In this article, we present a design approach for a three-dimensional user interface using 3D gestural input with the aim to facilitate user interaction within the context of Immersive Analytics.
Based on a scenario of exploring time-oriented data in immersive virtual reality using 3D Radar Charts, we implemented a rich set of features that is closely aligned with relevant 3D interaction techniques, data analysis tasks, and aspects of hand posture comfort.
We conducted an empirical evaluation (n=12), featuring a series of representative tasks to evaluate the developed user interface design prototype.
The results, based on questionnaires, observations, and interviews, indicate good usability and an engaging user experience.
We are able to reflect on the implemented hand-based grasping and gestural command techniques, identifying aspects for improvement in regard to hand detection and precision as well as emphasizing a prototype's ability to infer user intent for better prevention of unintentional gestures.
\end{abstract}

\begin{keyword}
empirical study\sep
immersive analytics\sep
user interface design\sep
virtual reality\sep
3D gestural input\sep
3D radar chart
\end{keyword}

\end{frontmatter}

\pagestyle{fancy} 
\lhead{\footnotesize \emph{Preprint submitted to arXiv.org cs.HC}} \chead{}\rhead{\footnotesize \emph{\today}}
\lfoot{}\cfoot{\footnotesize \thepage\ of \pageref*{LastPage}}\rfoot{}
\renewcommand{\headrulewidth}{0pt}

\section{Introduction}\label{sec:introduction}

Recent advances in immersive display and interaction technologies, such as head-mounted displays (HMD) and three-dimensional (3D) tracking sensors, have led to renewed interest in various research areas, especially outside entertainment-related contexts.
Immersive Analytics (IA), concerned with the application of immersive technologies for the purpose of data exploration, analysis, and meaning-making, is one such research area \citep{skarbez2019iat,dwyer2018iai}.
Among others, the utilization of immersive technologies for data analysis has the potential to increase user engagement \citep{buschel2018ifi}, promote user mobility \citep{fruchard2019lbi}, allow for the exploration of new data interaction approaches \citep{roberts2014vbt}, and enable the creation of virtual 3D data spaces to support collaborative decision making \citep{hackathorn2016iab}.
Within that context, the actual visualization of data in the Virtual Environment (VE) is arguably just one important aspect.
Equal importance in such immersive spaces should be attributed to their interactive features, enabling an analyst to actively explore and manipulate the VE.
Based on a recent survey, covering IA research from 1991 to 2018, \citet{fonnet2021soi} describe and discuss various aspects of such data interactions, highlighting the need for more guidelines and best practices as well as encouraging researchers to go beyond just basic interactions. 
The importance of interaction in IA systems has also been highlighted by \citet{ens2021gci}, deeming it a major topic within current IA research challenges.
Several studies across various contexts have shown that aspects of 3D gestural input for the interaction with immersive data visualizations can be generally intuitive, engaging, and easy to learn \citep{huang2017ags,wagner2020eai,reski2020eot}.
Nevertheless, there is still a need for further investigations, for instance to more clearly determine what types of 3D gestural interactions users prefer \citep{fittkau2015esc,streppel2018iiv}, or what kind of preferred user interactions are feasible to implement depending on current tracking capabilities \citep{austin2020esi}.

This article aims to address some of the research challenges by reporting on the design, implementation, and evaluation of a user interface that aims to facilitate engaging interaction with abstract data visualization in an immersive VE based on 3D gestural input and HMD technologies.
In particular, our research contributes to the emerging field of interactive IA as follows:
\begin{itemize}
    \setlength\itemsep{0em}
    \item We report on the design of a 3D user interface (3D UI) based on 3D gestural input with a focus on hand-based grasping and gestural command techniques, aimed to allow for engaging hand interaction with time-oriented data in immersive Virtual Reality (VR).
    \item We present an applied use case of mapping 3D interaction techniques, data analysis tasks, and aspects of hand posture comfort to the designed 3D UI, following an interdisciplinary research approach that informed and guided its 3D UI design.
    \item We present and discuss the results of an empirical evaluation of the developed 3D UI, allowing for reflections and considerations for similar future applications.
\end{itemize}

The structure of the article is such that Section~\ref{sec:relatedwork} provides theoretical insights in regard to 3D interaction techniques and data analysis task classifications as well as related work that features aspects of using 3D gestural input for immersive data interaction.
We present the developed 3D UI throughout Section~\ref{sec:vrprototypeand3duidesign}, describing the overall context and scenario, the adopted data analysis task terminology, the interface design and all its features, as well as details on technologies and implementation.
The methodology of our empirical evaluation is described in Section~\ref{sec:methodology}.
The results are presented in Section~\ref{sec:results} and discussed in Section~\ref{sec:discussion}.
The article is concluded in Section~\ref{sec:conclusion} with a brief summary and directions for future work.

\section{Related Work}\label{sec:relatedwork}

The presented context is concerned with the utilization of 3D gestural input as a means for interaction to perform analytical tasks in immersive VEs.
Therefore, Section~\ref{sec:relatedwork_3dinteractiontechniques} summarizes important foundational aspects in regard to 3D interaction techniques with a focus on 3D gestural input, i.e., hand interaction.
A brief overview of different data analysis task classifications is provided in Section~\ref{sec:relatedwork_dataanalysistaskclassifications}.
Thereafter, relevant related work that features 3D gestural input for interaction with abstract data in immersive VEs is discussed in Section~\ref{sec:relatedwork_immersivedatainteraction}.

\subsection{3D Interaction Techniques}\label{sec:relatedwork_3dinteractiontechniques}

An extensive overview of 3D interaction techniques is provided by \citet[Chapters~7--9]{laviola20173ui}, describing approaches and metaphors for \textit{selection and manipulation}, \textit{travel}, and \textit{system control} interaction techniques.
Particularly relevant in regard to 3D gestural input are \textit{grasping metaphors} \citep[Chapter~7]{laviola20173ui} that allow a user to simply \textit{grab, move, and release} a virtual artifact as one would do in the real world, as well as \textit{gestural commands} \citep[Chapter~9]{laviola20173ui} that utilize hand postures (static) and gestures (moving) that are associated with features to control the state of the VE.
It is important to differentiate between \textit{direct} and \textit{indirect} interactions: While \textit{direct} interaction allows for immediate manipulation of an object, \textit{indirect} interactions build upon some sort of middle layer for object manipulation, e.g., a representative proxy object or a virtual control widget \citep[Chapter~7]{laviola20173ui}.
Arguably, \textit{direct} interactions tend to be perceived as somewhat more natural than \textit{indirect} ones, as they reflect more closely how humans interact in the real world.
However, this does not mean that \textit{indirect} interactions should be avoided.
After all, interactions should aim firstly to be useful with regard to their intended purpose \citep{norman2010nui}.

Modern tracking sensors allow for interaction not just with one controller or hand but two \citep{bachmann2018rot}, commonly described as \textit{bimanual} metaphors \citep[Chapter~7]{laviola20173ui} that can be categorized with respect to their symmetry and synchronicity \citep{ulinski2009spb}.
\citet{pavlovic1997vio} reviewed aspects of 3D gestural input for application in Human-Computer Interaction (HCI) in general, describing a gestural taxonomy that (1)~differentiates hands and arm movements as \textit{gestures} and \textit{unintentional movements}, and (2)~divides \textit{gestures} into \textit{communicative} and \textit{manipulative} modalities.
Beyond hand and gesture recognition as fundamental prerequisites for any 3D gestural input \citep{pavlovic1997vio}, a computing system's ability to successfully infer intent in regard to subsequent hand interaction is equally important \citep{nehaniv2005ama}.
For instance, under consideration of the respective in-situ context, similar hand postures and gestures may be used for different types of interactions \citep{nehaniv2005ama}.
As such, \citet{nehaniv2005ama} classified gestures to infer human intent as \textit{irrelevant and manipulative gestures}, \textit{gestures as a side effect of expressive behaviour}, \textit{symbolic gestures}, \textit{interactional gestures}, and \textit{referential and pointing gestures}.
\citet{rempel2014tdo} provide considerations for the design of comfortable hand postures for the utilization in HCI contexts based on insights from sign language, among others to prevent physical fatigue symptoms.
The authors recommend the use of comfortable gestures for more frequent tasks, while infrequent tasks may also be performed through slightly less comfortable ones \citep{rempel2014tdo}.

\subsection{Data Analysis Task Classifications}\label{sec:relatedwork_dataanalysistaskclassifications}

Each visualization should be designed to serve a specific purpose and to accommodate the analyst with the extraction of insights and information by completing desired tasks.
\citet[Chapter~1.1]{aigner2011vot} summarize considerations for the design of information visualizations on a high level with respect to
(1)~\textit{what kind of data are visualized},
(2)~\textit{why are the data visualized}, and
(3)~\textit{how are the data going to be visualized}.
From a user-centred perspective, the specification of the analyst's tasks when interacting with a visualization is particularly interesting, i.e., with respect to why the data are visualized and what purpose the visualization serve the analyst.
\citet[Chapter~1.8]{ward2015idv} and \citet[Chapter~1.1]{aigner2011vot} differentiate between three main purposes for the interaction with visualizations:

\begin{itemize}
    \setlength\itemsep{0em}

    \item \textit{Exploration} or \textit{Explorative Analysis}: The analyst utilizes the visualization and its interactive features to explore an unknown dataset, and extract first insights and relevant information with no hypotheses given (undirected search). 

    \item \textit{Confirmation} or \textit{Confirmative Analysis}: The analyst utilizes the visualization and its interactive features to confirm or reject given hypotheses about a dataset (directed search).
    
    \item \textit{Presentation of Analysis Results}: The analyst utilizes the visualization and its interactive features to convey and present their findings in the dataset, such as concepts or facts, to an audience.
\end{itemize}

With respect to the actual design of a visualization's interactive capabilities, Shneiderman's \citep{shneiderman1996teh} Visual Information-Seeking Mantra of \textit{overview first, zoom and filter, then details-on-demand} is arguably one of the most famous design guidelines.
Based on it, \citet{shneiderman1996teh} proposes seven abstract task types that should be supported by the visualization, namely \textit{overview}, \textit{zoom}, \textit{filter}, \textit{details-on-demand}, \textit{relate}, \textit{history}, and \textit{extract}.
Another approach by \citet[Chapters~1--3]{munzner2014vad} and \citet{brehmer2013aml}, describes abstract visualization tasks as a multi-level typology, organizing tasks as to \textit{why} and \textit{how} they are performed as well as \textit{what} a task's input and output parameters are.
With respect to why, \citet[Chapter~3]{munzner2014vad} classifies user actions across four overall groups, i.e.,
(1)~\textit{analyze} (\textit{discover}, \textit{present}, \textit{enjoy)},
(2)~\textit{produce} (\textit{annotate}, \textit{record}, \textit{derive)},
(3)~\textit{search} (\textit{lookup}, \textit{locate}, \textit{browse}, \textit{explore)}, and
(4)~\textit{query} (\textit{identify}, \textit{compare}, \textit{summarize}).
Depending on the scenario and context of the interactive visualization, all these classifications have the potential to be informative for the development, either in isolation or as a mixed and multimodal approach.
This allows for guidance and facilitation of the design process towards purposeful interactions with a visualization, and thus with data.
\citet{yi2007tad} reviewed a multitude of information visualization taxonomies with respect to their described interaction techniques.
Based on their analysis of the literature, they synthesized a set of formal categories (\textit{select}, \textit{explore}, \textit{reconfigure}, \textit{encode}, \textit{abstract/elaborate}, \textit{filter}, \textit{connect}, \textit{undo/redo}, \textit{change configuration}) to describe a user's intent for the interaction with a visualization in general \citep{yi2007tad}.
\citet[Chapter~5.1]{aigner2011vot} further build upon these categories and adapt them to support the more specific context of interacting with \textit{time-oriented} data, i.e., multivariate data where each data item features at least one data variable related to a temporal context.
The utilization of such task categories allows us to conceptually categorize the interactive features of a developed data analysis tool, similar as presented by \citet{buschel2018ifi}, and thus aiding the tool's description accordingly.

\subsection{3D Gestural Input for Immersive Data Interaction}\label{sec:relatedwork_immersivedatainteraction}

\citet{laviola2000mav} describes an interface that utilizes a multimodal approach of 3D gestural input and voice commands to interact with a scientific data visualization in stereoscopic 3D.
Different analysis tools can be attached to the user's hands and moved around in the 3D environment. 
Interestingly, rather than selecting these tools from a graphical menu, they implemented voice commands that allow the user to say aloud the tool they want to interact with, following a ``\textit{show and ask}'' metaphor. 
They also implemented several hand-based grasping configurations to provide navigation features, i.e., user movement as well as translation, scaling, and rotation of the VE. 
An evaluation indicated that their participants valued the tool's ease of use after an initial learning phase. 
Their results also indicated that the voice command interface worked well in single-user scenarios, while having detection problems in collaborative ones that featured more auditory input.

\citet{fittkau2015esc} explored gestural command design for interaction with an immersive data visualization following the ``\textit{software cities}'' metaphor, implementing several unimanual and bimanual hand commands to support translation, rotation, zoom, selection, and reset tasks.
The results of their evaluation indicate that the users favoured one-handed gestures (translation, rotation, selection) over the two-handed (zoom) one that was performed through a ``\textit{rowing}'' motion. 
Interestingly, the authors attempted to utilize more elements of embodied interaction for the zooming command, such as rotating the user's torso or walking back and forth in the VE. 
However, such movements would inherently result in a change of the user's point of view, which where not appreciated during early design iterations. 

\citet{streppel2018iiv} explored 3D interaction techniques within the ``\textit{software cities}'' context as well, comparing 3D gestural input, physical controllers, and virtual controls.
Their results indicate similar preferences for 3D gestural input and physical controllers as opposed to virtual controls. 
Even though the physical controller condition received better usability scores, participants stated that they would rather like to use the 3D gestural input in a real-world scenario, as it was perceived as more natural and appropriate for interaction in a VE. 
The expressed desire for better 3D gestural input controls is quite interesting, indicating that more work in that direction should be undertaken to further improve the usability aspects of 3D gestural input in the context of IA.

\citet{osawa2000ign} investigated hand-based grasping and gestural command techniques for interaction with an immersive graph visualization.
Their system allowed the user to select and manipulate individual nodes of the 3D network (translate, lock position in space, adjust characteristics), to translate the user's position in space (move), and to adjust characteristics of multiple nodes through a ``\textit{spotlight}'' approach. 
The latter was operated by pointing one's hand in the general direction of the desired nodes, and creating an arc-like spread by moving the index finger and thumb apart, enabling dynamic control of the included network nodes. 

\citet{huang2017ags} reported on the design of a 3D gestural interface for interaction with graph visualizations in VR, providing gestures to move and highlight nodes and edges (one-handed interaction), to rotate and translate the entire graph, and to group nodes (two-handed interaction).
An evaluation, comparing the implemented gestures with more traditional pointer input (mouse), revealed positive trends towards the participants' ability to manipulate the 3D graph with the gestures, stating that the interface ``\textit{was intuitive, easy to learn, and interesting}''. 
While their implemented node/edge movement and graph rotation gestures were appreciated for their learnability, some usability issues were identified for the highlight and group gestures that involved aspects such as holding a specific hand posture or performing a gesture very quickly. 

A VR system developed by \citet{betella2014uln} featured 3D gestural input for manipulation and filter operations within a large network visualization.
Their interface utilized a hand-based grasping technique and asymmetric bimanual hand interaction, i.e., one of the user's hands had a cursor function to highlight and select elements in the network, while the other hand was used to operate task parameters such as filter strength and complexity. 
Their asymmetric feature mapping strategy is interesting insofar that the authors differentiate between left and right-hand interactions instead of following a symmetric approach where the same features are provided independently of which hand performs the posture or gesture.

As part of interacting with an immersive 3D trajectory visualization, \citet{wagner2020eai} implemented a mixture of hand-based grasping (scale, translate) and gestural commands (single and double tap via index finger to inspect and select).
They evaluated their system in comparison to a desktop one, revealing generally better usability scores for the immersive VE. 
Participants overall agreed that the 3D gestural input enabled them to easily and comfortably manipulate the data, resulting in an engaging and intuitive experience.
Room for improvement was identified towards the index finger tapping that required to be comparatively precise, and towards the two-handed scale and rotation commands whose similar operation was sometimes perceived as too constraining. 

\citet{austin2020esi} investigated common hand gestures for the interaction with large immersive maps that are placed on a virtual floor.
In particular, using a participatory design approach, their study participants were asked to come up with hand gestures for typical operations to manipulate the virtual map, such as pan, rotate, zoom, and marker interaction. 
Their results indicate that the participants most commonly proposed unimanual gestures for interactions such as pan as well as creating and selecting markers on the map, while proposing bimanual gestures for rotate and zoom operations. 
\citet{austin2020esi} reflect on their findings, stating that identified user preferences for these gestural commands need further investigation in regard to performance-related matters, such as efficiency, accuracy, and physical fatigue. 
\citet{austin2020esi} also reflect on potential feasibility concerns for some of the proposed gestural commands, stating that an accurate and reliable implementation based on current 3D gestural tracking sensors might be difficult.

\section{3D UI Design and VR Prototype}\label{sec:vrprototypeand3duidesign}

As seen throughout Section~\ref{sec:relatedwork}, there is a multitude of aspects worth considering when setting out to design a 3D UI for immersive data interaction.
We begin by describing details about the context and scenario in Section~\ref{sec:vrprototype_context}, providing an entry point for our VR prototype.
Section~\ref{sec:vrprototype_adopteddataanalysistaskterminology} presents the data analysis task terminology that we adopted for the immersive interaction with spatio-temporal data.
Design and motivation for the developed 3D gestural interface are described in Section~\ref{sec:vrprototype_design}, presenting an overview of all features with respect to relevant taxonomies.
A brief summary of involved technologies and implementation is provided in Section~\ref{sec:vrprototype_implementation}.

\subsection{Context and Scenario}\label{sec:vrprototype_context}

\begin{figure}
        \begin{center}
        \includegraphics[width=.9\columnwidth]{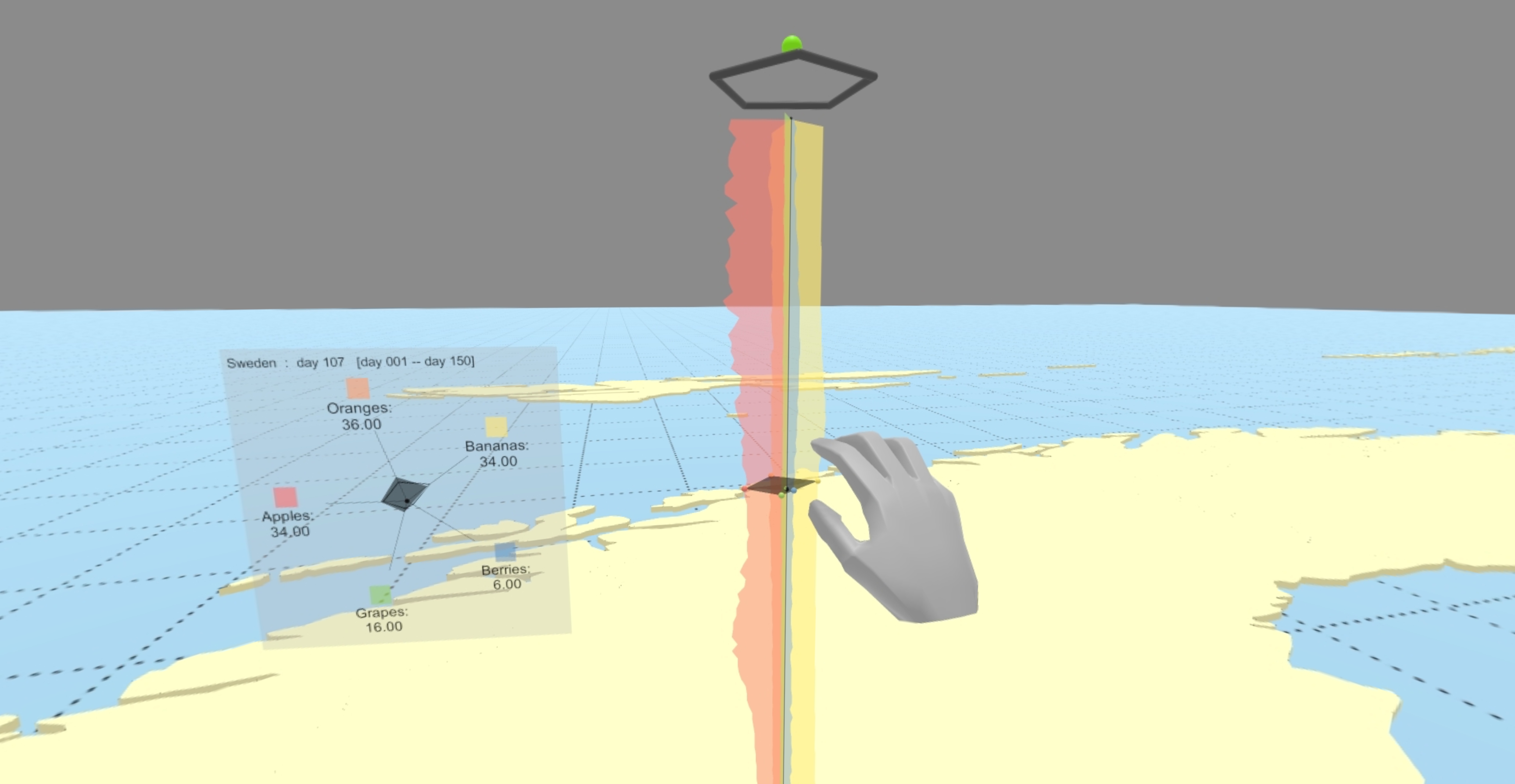}
        \end{center}
    \caption{Excerpt of a VE from the VR user's field of view, and interacting through 3D gestural input with a 3D Radar Chart (incl. Time Slice and juxtaposed Information Window).}\label{fig:3dradarchart}
\end{figure}

The focus of this article is to investigate 3D UI design to support user interaction with abstract data visualization using 3D gestural input (hand interaction) within the context of IA.
More specifically, we are interested in the interaction with time-oriented data in immersive spaces, a comparatively common IA use case \citep{fonnet2021soi}.
For this purpose, we build upon the 3D Radar Chart approach as presented by \citet{reski2020eot}.
Their approach allows for time-oriented data visualization in immersive VR, enabling the user to explore multivariate data in regard to spatial and temporal dimensions. 
Conceptually, a 3D Radar Chart consists of a central Time Axis with multiple Data Variable Axes arranged radially around it, each depicting a respective time-series visualization. 
A two-dimensional interactive Time Slice, illustrating the more traditional radar chart-like pattern \citep{kolence1973sup}, allows for temporal analysis of the values across the different data variables \citep{reski2020eot}.
A VE may be populated with multiple 3D Radar Chart instances, each representing a different entity in the data, e.g., a location, thus allowing for spatio-temporal data analysis. 
\autoref{fig:3dradarchart} presents the described concept of a 3D Radar Chart, providing an excerpt of a VE from the VR user's field of view.

The results of their initial study validated the visualization approach in general, indicating that the participants were able to explore and interpret the displayed time-series data using a first set of basic interaction features, such as selecting time events and time ranges. 
As part of their initial explorative interaction design, \citet{reski2020eot} implemented alternatives for the interaction using hand-based grasping as well as system control (via graphical menus attached to the user's virtual hand) techniques.
However, no clear preference for one interaction technique over the other could be identified by \citet{reski2020eot}. 
Compared to the initial prototype in \citep{reski2020eot}, we rigorously iterated on the 3D UI design in several aspects as follows:
\begin{enumerate}
    \setlength\itemsep{0em}
    \item Beyond the basic set of interactions, we extended the prototype with new features to support additional important tasks that are typical for the analysis of time-oriented data, such as sort, filter, and zoom capabilities \citep{yi2007tad,aigner2011vot}.
    \item Within the design and implementation of the 3D gestural input we focus on hand-based grasping and gestural command techniques with the objective to provide a uniform interaction approach, i.e., without the use of any alternative graphical menu-based system control techniques.
    \item We implemented quality-of-life features to further enhance the 3D Radar Chart approach in general. For instance, we provide a semi-transparent uncolored preview of the data outside the selected time range, instead of simply hiding the unselected data, facilitating the user's time range selection experience.
\end{enumerate}

The implemented changes and extensions allow for the contribution to the IA community with further insights towards the applied design of 3D gestural input for the interaction with time-oriented data in immersive VR.

\subsection{Data Analysis Tasks for Immersive Interaction With Spatio-Temporal Data}\label{sec:vrprototype_adopteddataanalysistaskterminology}
Based on the combined work presented by \citet{yi2007tad} and \citet[Chapter~5.1]{aigner2011vot}, we adopted their data analysis tasks towards the contexts of IA and the interaction with spatio-temporal data in VEs as follows:

\begin{enumerate}
    \setlength\itemsep{0em}

    \item \textit{Select -- Mark something as interesting}:
    Select a data entity at a specific spatial location in the VE \underline{or} modify the displayed temporal context through the selection of a new time event or time range, for instance with the objective to perform various follow-up interactions, such as to display details-on-demand.
    
    \item \textit{Explore -- Show me something else}:
    Look around in the VE with the objective to identify a location/region (spatial) or time event/range (temporal) of interest worthy of further inspection \underline{or} move around in the VE in order to reach data entities, either in close proximity or far away (outside the physical real-world boundaries of the VR system's calibrated safe interaction area), potentially utilizing virtual travel features.
    
    \item \textit{Reconfigure -- Show me a different arrangement}:
    Perform an interaction that modifies the visual arrangement of the displayed data entities in the VE, for instance with respect to their relative location in the VE or in regard to aspects of their individual visual representation (for instance, sorting the order of the displayed data variables).
    
    \item \textit{Encode -- Show me a different representation}:
    Modify the visualization technique used to represent a data entity in the VE, i.e., mapping a data item's data variables onto a new visual representation and in turn creating a different data entity.
    
    \item \textit{Abstract/Elaborate -- Show me more or less detail}:
    Aligned with Shneiderman's Visual Information-Seeking Mantra \citep{shneiderman1996teh}, display details-on-demand (elaborate) to show additional information about a selected data entity, or hide the details (abstract) to enable a more overview-like perspective and interaction mode.
    
    \item \textit{Filter -- Show me something conditionally}:
    Perform an interaction that modifies the visual representation of one or more data entities in the VE to conditionally hide or add information, for instance by deactivating entire data entities or aspects of their individual visual representation (for instance, filtering out undesired displayed data variables).
    
    \item \textit{Connect -- Show me related items}:
    Perform an interaction in the VE that facilitates the inference of relationships between and the comparison of data entities, both with respect to spatial and temporal contexts.
    
    \item \textit{Undo/Redo -- Let me go to where I have already been}:
    With respect to the interaction in the VE in general, enable the user to retrace their previous interactions, for instance through undo, redo, history, or reset functionalities.
    
    \item \textit{Change Configuration -- Let me adjust the interface}:
    Perform an interaction that modifies aspects of the user interface on a system level in general \underline{or} with respect to the particular in-situ interaction mode with one or multiple selected data entities (for instance, temporally accessing and switching between menus and widgets that assist with the interaction in the VE).
\end{enumerate}

\subsection{3D Gestural Interface Design}\label{sec:vrprototype_design}

\begin{figure}
    \begin{center}
    \includegraphics[width=.9\columnwidth]{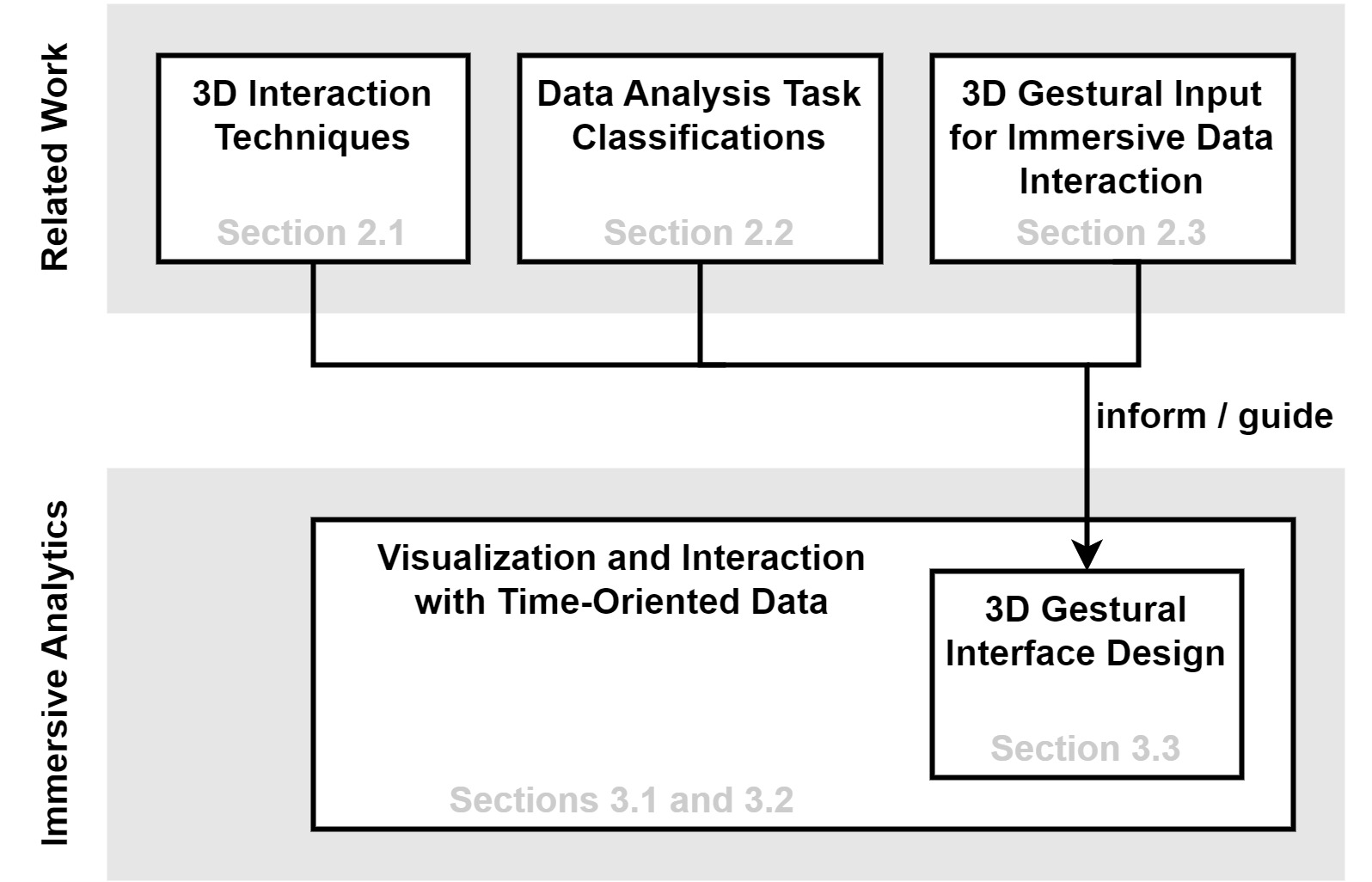}
    \end{center}
    \caption{Illustration of the 3D UI design approach.}\label{fig:3duidesignapproach}
\end{figure}

Following a prototypical approach, we designed a 3D gestural interface for the interaction with 3D Radar Charts in VR under consideration of various theoretical aspects, practical guidelines, recommendations, and lessons learned from related work as described throughout Section~\ref{sec:relatedwork}.
\autoref{fig:3duidesignapproach} illustrates the overall design approach.
We started with an overall task analysis, aiming to identify the particular interactions an analyst would likely perform when exploring time-series data.
For this purpose, we adopted the data analysis tasks as described throughout  Section~\ref{sec:vrprototype_adopteddataanalysistaskterminology}.
The actual design of the interaction was conceptually informed by the various 3D interaction technique classifications according to \citet[Chapters~7--9]{laviola20173ui}, with additional considerations in regard to hand posture comfort as discussed by \citet{rempel2014tdo}.

We envision \textit{explorative analysis} \citep[Chapter~2]{aigner2011vot} as one of the main use cases for such interaction with time-series data, i.e., using the immersive VE for data explorations and observations to extract first insights that can lead to subsequent analysis.
As such, the analyst is arguably going to perform certain task types more frequently than others. 
This requires keeping in mind hand comfort recommendations such as those reported by \citet{rempel2014tdo} in order to avoid the use of uncomfortable hand configurations for anticipated frequent interactions.
Under the assumption that the VE is populated with a variety of data entities (3D Radar Charts), each representing different time-series data, means for spatial exploration are needed, i.e., a \textit{Travel} feature to enable movement in the VE beyond physical space limitations.
This allows for utilization of the virtual 3D space by instantiating many data entities, enabling the user to explore the data in a more \textit{overview}-like manner \citep{shneiderman1996teh}, conceptually similar to ``\textit{walking among the data}'' \citep{ivanov2019awa,streppel2018iiv}.
Naturally, when discovering something of interest, the user is expected to engage in-situ with the data to display \textit{details-on-demand} \citep{shneiderman1996teh}, thus entering a closer contextual interaction \citep{nehaniv2005ama}.
At this stage, we can expect the user to
(1)~\textit{Select Time Events and Time Ranges} as well as potentially (2)~\textit{Reconfigure (Sort)} the order and
(3)~\textit{Filter} out individual data variables.

Besides these envisioned frequent tasks, we also considered features for more infrequent ones.
Depending on the number of time events in the time series as encoded over the static length of a 3D Radar Chart (its height in the VE), we were interested in providing a \textit{Zoom} feature.
With a time range selected, the user may \textit{Zoom in} by temporarily ``stretching'' their time-series selection over the entire virtual length of the 3D Radar Chart, visually cutting off any time events outside that range.
Reversely, assuming the entire time series is not already displayed, the user may also \textit{Zoom out} from previous \textit{Zoom in} interactions.
We implemented a history feature, allowing for step-wise \textit{Zoom out} based on multiple prior \textit{Zoom in} interactions.
It is also important to provide the user with means to reverse selections and manipulations, and therefore implemented a \textit{Reset} feature, conveniently reconfiguring a 3D Radar Chart back to its original state.
Finally, we wanted to explore the possibility to allow the user to temporary \textit{pause} any kind of interaction, for instance, to avoid unintentional hand movements \citep{pavlovic1997vio} during periods when the user desires to make observations in the VE more passively.

\begin{figure}
        \begin{center}
            \includegraphics[width=.9\columnwidth]{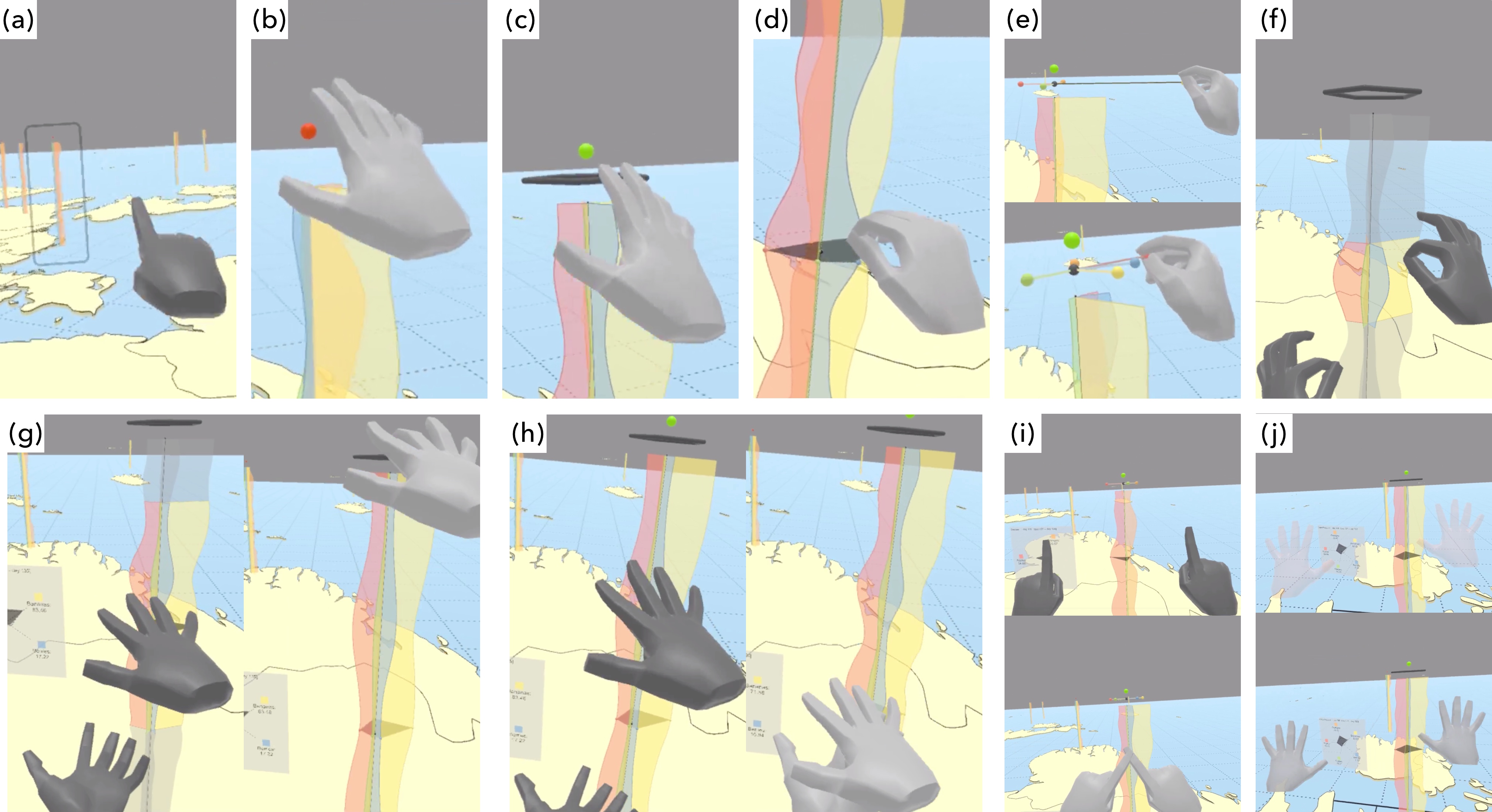}
        \end{center}
    \caption{Feature overview of the implemented 3D gestural interface (see \autoref{tab:3duioverview}):
    (a)~Travel;
    (b)~Mode Toggle;
    (c)~Rotation;
    (d)~Time Event Selection;
    (e)~Data Variable Filter (top) and Sort (bottom);
    (f)~Time Range Selection;
    (g)~Zoom (in);
    (h)~Zoom (out);
    (i)~Reset;
    (j)~Pause (top) / Resume (bottom).
    See also video demonstration (URL provided in Footnote \ref{footnote:demo}).
    }\label{fig:3duioverview}
\end{figure}

\begin{figure}
    \begin{center}
    \includegraphics[width=.9\columnwidth]{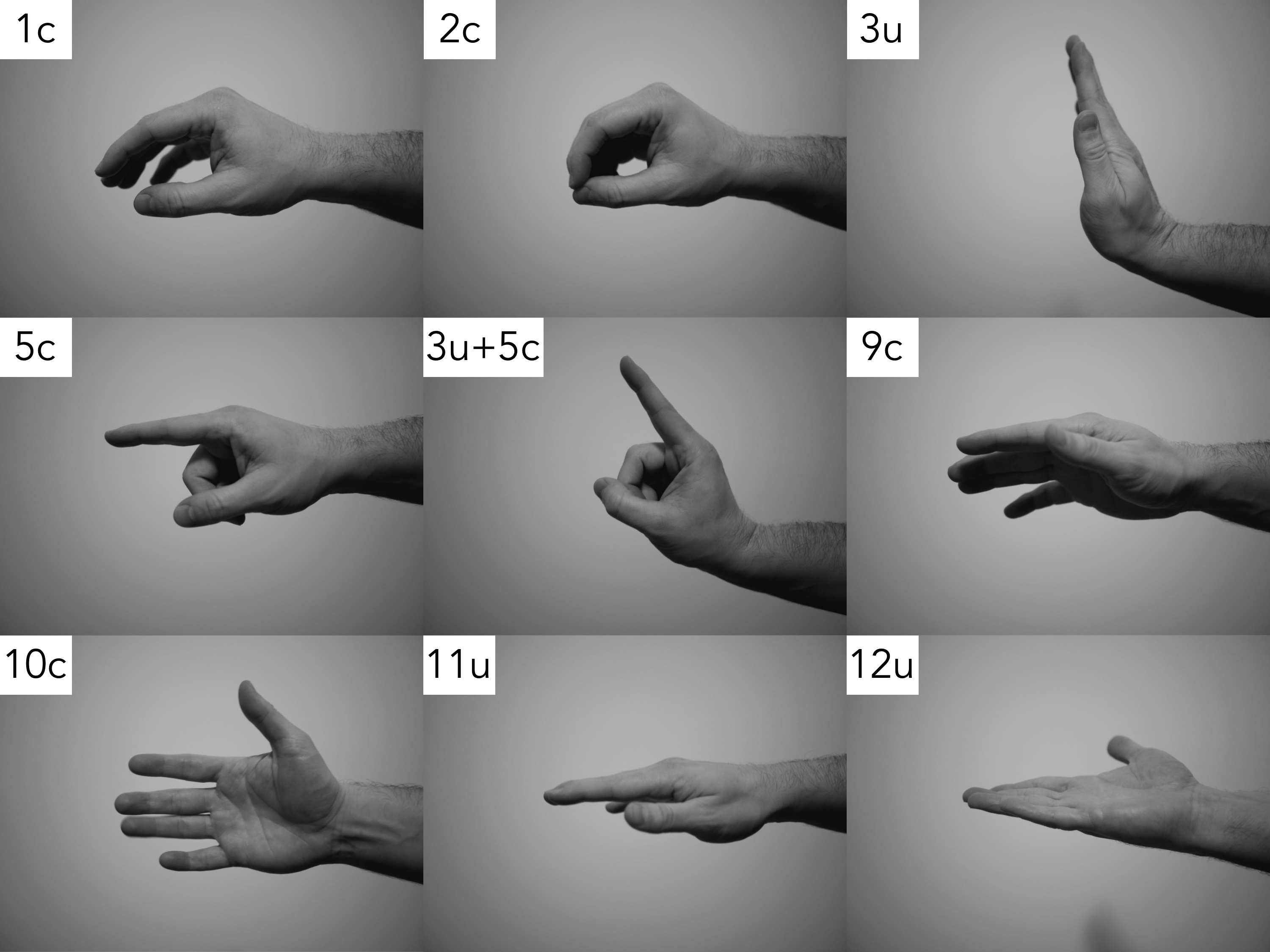}
    \end{center}
    \caption{Applied hand posture comfort configurations, adapted from the recommendations by \citet{rempel2014tdo}. To facilitate cross-referencing to their work \citep[Figure~5]{rempel2014tdo}, we apply here the same label coding for convenience: 1--12 = hand posture example; c = comfortable; u = uncomfortable.}\label{fig:handpostures}
\end{figure}

To support these anticipated tasks and interactions, we designed the 3D gestural interface with a focus on hand-based grasping interaction with virtual objects as well as through means of gestural commands based on the user's in-situ context.
Based on our interest and within the scope of this investigation, we deliberately avoided graphical menu-based system control techniques \citep[Chapter~9]{laviola20173ui}.
We kept hand posture comfort recommendations in mind \citep{rempel2014tdo}, prioritizing seemingly more comfortable hand postures for anticipated frequent tasks in the VE.
\autoref{fig:3duioverview} demonstrates the 3D gestural interface in the immersive VR environment.\footnote{\label{footnote:demo}Video demonstration of the developed 3D gestural interface (3:57~min, no~audio): \textcolor{cyan}{\href{https://vrxar.lnu.se/arxiv/reski2023dat.mp4}{vrxar.lnu.se/arxiv/reski2023dat.mp4}}}
\autoref{fig:handpostures} illustrates the real-world hand posture configurations we applied within the scope of the presented 3D gestural interface.
\autoref{tab:3duioverview} provides a comprehensive overview of all implemented features of the 3D gestural interface, including their data analysis task, interaction technique, and comfort classification.

\subsection{Technologies and Implementation}\label{sec:vrprototype_implementation}

The VR prototype utilizes an HTC Vive HMD (1080x1200 pixel resolution per eye, 90 Hz refresh rate) and a Leap Motion controller (10-80~cm interaction zone depth, 120x150\textdegree~field of view, Ultraleap Hand Tracking V4 Orion software, attached to the HMD's front) for the 3D gestural input.
Both devices are commercially available.
The HTC Vive is configured as room-scale VR with a 2x2~meter area for the user to move freely without any physical obstacles.
Unity~2019.3, SteamVR Plugin for Unity~1.2.3, and Leap Motion Core Assets~4.5.1 have been used to develop the prototype.

\section{Evaluation Methodology}\label{sec:methodology}

To assess the developed 3D UI design, we conducted an empirical evaluation using a series of representative tasks, questionnaires, and interviews \citep[Chapter~11]{laviola20173ui}.
Allowing human users to go hands-on with the prototype enables us to apply subjective methods to collect quantitative and qualitative data, and thus to evaluate its design.
This section describes the setup, task, applied measures, procedure as well as ethical considerations.

\begin{table}
  \caption{Summary of the 3D gestural interface design to interact with time-series data in the immersive VE as presented in \autoref{fig:3duioverview}, utilizing the 3D Radar Chart approach by \cite{reski2020eot}, including classifications in regard to
  time-oriented \textbf{data analysis tasks} (see Section~\ref{sec:vrprototype_adopteddataanalysistaskterminology}),
  3D \textbf{interaction techniques} \cite[Chapters~7~,8,~and~9]{laviola20173ui}, and
  hand posture \textbf{comfort} \cite[Figure~5]{rempel2014tdo} as illustrated in \autoref{fig:handpostures}.}\label{tab:3duioverview}
  \scriptsize
  \centering
  \begin{tabu}{@{}l@{\hspace{1ex}}llll@{}}
  \toprule
  \textbf{Feature} &
  \begin{minipage}{4.9cm} \textbf{Interaction Description} \end{minipage} &
  \textbf{Task} &
  \begin{minipage}{29ex} \textbf{Interaction Technique} \end{minipage} &
  \textbf{Comfort} \\
  
  \midrule
  Travel &
  \begin{minipage}{4.9cm} Look at a faraway 3D Radar Chart until its outline is displayed, then point towards it (left/right hand index finger pointing forward) to initiate position transition via target-based travel. \end{minipage} &
  Explore &
  \begin{minipage}{29ex} Selection-based Travel \\ (Multimodal Technique: \\Gaze-based Input and \\Gestural Command)\end{minipage} &
  5c \\
  
  \midrule
   \begin{minipage}{1cm} Mode Toggle \end{minipage} &
  \begin{minipage}{4.9cm} Touch virtual \textit{Mode Toggle} widget to iterate between three states: (1) Activate/Rotate, (2) Reconfigure/Filter, and (3) Deactivate. \end{minipage} &
  \begin{minipage}{1.25cm} Abstract\\/Elaborate, \\ Change\\ Configuration \end{minipage} &
  \begin{minipage}{29ex} Hand-Based Grasping \end{minipage} &
  \begin{minipage}{1.3cm} 1c, 5c, 9c \\ (either) \end{minipage}\\

  \midrule
  Rotation &
  \begin{minipage}{4.9cm} Grab \textit{Rotation Handle} widget and drag it left/right to rotate around its Time Axis \underline{or} give the \textit{Rotation Handle} widget a little left/right flick with the whole hand. \end{minipage} &
  \begin{minipage}{1.25cm} Change \\ Configuration \end{minipage} &
  \begin{minipage}{29ex} Indirect Widget \\ (Hand-Based Grasping) \end{minipage} &
  \begin{minipage}{1.3cm} 2c, 10c \\ (either) \end{minipage}\\
  
  \midrule
  \begin{minipage}{1cm} Data \\ Variable \\ Sort \end{minipage} &
  \begin{minipage}{4.9cm} Grab \textit{Data Variable Axis Sphere} widget, drag it around the Time Axis, and release it to apply the new radial arrangement. \end{minipage} &
  Reconfigure &
  \begin{minipage}{29ex} Indirect Widget \\ (Hand-Based Grasping)\end{minipage} &
  2c \\
  
  \midrule
  \begin{minipage}{1cm} Data \\ Variable \\ Filter \end{minipage} &
  \begin{minipage}{4.9cm} Grab \textit{Data Variable Axis Sphere}, drag it away from the Time Axis until its visual connection disappears (``snaps''), and release it to remove the associated \textit{Data Variable Axis}. \end{minipage} &
  Filter &
  \begin{minipage}{29ex} Indirect Widget \\ (Hand-Based Grasping) \end{minipage} &
  2c \\
  
  \midrule
  \begin{minipage}{1cm} Time Event \\ Selection \end{minipage} &
  \begin{minipage}{4.9cm} Grab \textit{Time Slice}, and drag it up (forward in time) or down (backwards in time) to select a new time event. \end{minipage} &
  Select &
  \begin{minipage}{29ex} Hand-Based Grasping \end{minipage} &
  2c \\
  
  \midrule
  \begin{minipage}{1cm} Time Range \\ Selection \end{minipage} &
  \begin{minipage}{4.9cm} Pinch (index finger and thumb held together) with each hand to unfold a highlighted time range. As long as the hands remain in that posture, the selected time range is updated, allowing to move the hands closer together/further apart for preview. Releasing the pinch applies the time range selection. \end{minipage} &
  Select &
  \begin{minipage}{29ex} Symmetric Bimanual \\ (Gestural Command)\end{minipage} &
  \begin{minipage}{1.3cm} 2c + 2c \\ (together) \end{minipage}\\ 
  
  \midrule
  Zoom in &
  \begin{minipage}{4.9cm} With a time range selected, hold both hands with their palms facing each other, and move them apart, ``stretching'' the selected time range over the entire length of the 3D Radar Chart. \end{minipage} &
  Elaborate &
  \begin{minipage}{29ex} Symmetric Bimanual \\ (Gestural Command)\end{minipage} &
  \begin{minipage}{1.3cm} 11u + 12u \\ (together) \end{minipage}\\
  
  \midrule
  Zoom out &
  \begin{minipage}{4.9cm} Hold both hands with their palms facing each other, and move them towards each other (``clapping'') to apply the previous time range over the entire length of the 3D Radar Chart. \end{minipage} &
  Abstract &
  \begin{minipage}{29ex} Symmetric Bimanual \\ (Gestural Command)\end{minipage} &
  \begin{minipage}{1.3cm} 11u + 12u \\ (together) \end{minipage}\\
  
  \midrule
  Reset &
  \begin{minipage}{4.9cm} Hold both hands with index fingers pointing upwards, then move index fingers to cross each other (``X''-like posture) to reset the state of the 3D Radar Chart
  (display entire time series and all data variables in original arrangement).
  \end{minipage} &
  Undo &
  \begin{minipage}{29ex} Symmetric Bimanual \\ (Gestural Command)\end{minipage} &
  \begin{minipage}{1.3cm} 5c and 3u \\ (composite) \end{minipage}\\
  
  \midrule
  \begin{minipage}{1cm} Pause/ \\ Resume \end{minipage} &
  \begin{minipage}{4.9cm} Hold both hands stretched out in front of the torso in a ``stop''-like posture for 1.5~seconds to iterate between two states: (1)~Paused (hands semitransparent, no interactions available), and (2)~Resumed (hands opaque, all interactions available). \end{minipage} &
  \begin{minipage}{1.25cm} Change \\ Configuration \end{minipage} &
  \begin{minipage}{29ex} Symmetric Bimanual \\ (Gestural Command)\end{minipage} &
  \begin{minipage}{1.3cm} 3u + 3u \\ (together) \end{minipage}\\
  
  \bottomrule
  \end{tabu}
\end{table}

\subsection{Physical Study Space and Virtual Environment}

Each study session involved one participant and one researcher, who was moderating the study, collecting data, and ensuring that all hard- and software components were working as intended.
Our research group lab provided enough space for both to conduct the study, including a dedicated space for the VR user, the researcher's workstation, several chairs, and a participant desk that was physically partitioned from the researcher's workstation.
The researcher remained at their workstation at all times for the study moderation (introduction, prototype initialization, tasks) and data collection (observation, note taking, interview).
The participant was seated twice at their desk to complete the informed user consent form (pre-task) and questionnaires (post-task), while otherwise remaining in the VR area (tasks) and its adjacent chairs (post-task interview).

We set up the VE as a representative IA scenario to allow for spatio-temporal data exploration as follows.
European countries are displayed as extruded polygons on the floor.
The VE is populated with 39 3D Radar Charts, each respectively placed at the center of a country.
Each 3D Radar Chart features five data variables, each with a time series of 150~consecutive time events (\textit{per day} basis).
We artificially generated all the time-series data for this 3D UI design evaluation.
The data scenario was conceptually designed to be approachable, demanding no specific prior knowledge, allowing for inclusive participant recruitment with no expert requirements.
The five data variables were labeled as various types of \textit{fruits} (Apples, Oranges, Bananas, Berries, Grapes), representing \textit{fruit production over time}.
This scenario allows for spatial (European countries) and temporal (time series at each country) data exploration featuring an easily understandable data context.
All implemented features were available to the VR user (see Section~\ref{sec:vrprototype_design}).
They could freely move within the physical space and interact with 3D Radar Charts in close proximity, or \textit{Travel} to virtually distant locations.

\subsection{Tasks}\label{sec:task}

We created a series of 31 tasks (see~\autoref{tab:taskseries}), comprising a mixture of all implemented features, and structured to be representative of a typical analytical session, using the prototype in a walkthrough-like manner.
We included \textit{definite} tasks (e.g., navigate to time event~X) as well as \textit{indefinite} tasks (e.g., select the event~X you deem appropriate), enabling participants to partially make their own data observations and interpretations.
All participants started at the same location (the eastern border of all 3D Radar Charts). 
The researcher would read aloud each next task to the participant upon completion of the prior one. 
The same task series order was applied across all participants, and their spoken-aloud answers were noted by the researcher on a task answer sheet.

\begin{table}
    \caption{Series of tasks and their associated interaction feature (see \autoref{tab:3duioverview}), completed by each participant within the scope of the empirical evaluation. Annotations: $^*$ \textit{ensure understanding of visualization concept} (T06, T07); $^{**}$ \textit{interaction paused demonstration} (T26).}\label{tab:taskseries}
    \renewcommand{\arraystretch}{1.1}
    \centering
    \footnotesize
    \begin{tabu}{@{}lll@{}}
    \toprule
    \begin{minipage}{5mm} \textbf{No.} \end{minipage} &
    \begin{minipage}{70mm} \textbf{Task} \end{minipage} &
    \begin{minipage}{29mm} \textbf{Feature} \end{minipage} \\
    
    \midrule
    
    T01 &
    Move to \textit{Italy}.  &
    Travel \\
    
    \midrule

    T02 &
    Move to \textit{Sweden}. &
    Travel \\

    T03 &
    Activate the 3D Radar Chart at your current location.&
    Mode Toggle \\
    
    T04 &
    Navigate to day \textit{120}. &
    Time Event Selection \\

    T05 &
    Rotate the 3D Radar Chart entirely around its own~axis. &
    Rotation \\
    
    T06 &
    Name the data variable with the second highest value. &
    * \\
    
    T07 &
    Name the data variable with the second lowest value. &
    * \\
    
    T08  &
    Select a time range you deem appropriate that contains &
    Time Range Selection \\
     &  three peaks in the \textit{Berries} variable. & \\   
    
    T09 &
    Zoom in into the selected time range. &
    Zoom (in) \\
    
    T10 &
    Select a time range you deem appropriate that contains & 
    Time Range Selection \\
     & one valley in the \textit{Oranges} variable and one valley in the & \\ 
     & \textit{Grapes}~variable. & \\ 

    T11 &
    Zoom in into the selected time range. &
    Zoom in \\
    
    T12 &
    Zoom out once. &
    Zoom out \\
    
    T13 &
    Switch to the reconfigure and filter mode. &
    Mode Toggle \\
    
    T14 &
    Navigate to a time event of your choice that you deem &
    Time Event Selection \\
    &  interesting, and briefly describe why it is interesting to & \\ 
    &  you. & \\ 
    
    T15 &
    For the currently selected time event, sort all data  &
    Data Variable Sort \\
    & variables in ascending order based on their value. & \\ 
    
    T16 &
    Zoom out once. &
    Zoom out \\
    
    T17 &
    Reset the state of the 3D Radar Chart. &
    Reset \\
    
    T18 &
    Deactivate the 3D Radar Chart at your current location. &
    Mode Toggle \\
    
    \midrule
    
    T19 &
    Move to \textit{Italy}. &
    Travel \\
    
    T20 &
    Activate the 3D Radar Chart at your current location. &
    Mode Toggle \\
    
    T21  &
    Switch to the reconfigure and filter mode. &
    Mode Toggle \\
    
    T22 &
    Navigate to day \textit{56}.&
    Time Event Selection \\

    T23 &
    For the currently selected time event, remove all the data &
    Data Variable Filter \\
    & variables with a value lower than \textit{20}. & \\ 
    
    T24 &
    Reset the state of the 3D Radar Chart. &
    Reset \\
    
    T25 &
    Pause the 3D hand interaction. &
    Pause \\
    
    T26 &
    Attempt to navigate to a different time event.&
    **  \\
    
    T27 &
    Resume the 3D hand interaction. &
    Resume \\
    
    T28 &
    Navigate to a time event of your choice that you deem &
    Time Event Selection \\
    &  interesting, and briefly describe why it is interesting to you. & \\ 
    
    T29 &
    Navigate to day \textit{98}.&
    Time Event Selection \\
    
    T30 &
    For the currently selected time event, sort all data variables &
    Data Variable Sort \\
    &  in descending order based on their value. & \\ 
   
    T31 &
    Deactivate the 3D Radar Chart at your current location. &
    Mode Toggle \\
    
    \bottomrule
    \end{tabu}
\end{table}

\subsection{Quantitative and Qualitative Measures}\label{sec:measures}

To make a generalized assessment of the prototype's usability, we utilized the \textit{System Usability Scale} (SUS) questionnaire \citep{brooke2013sus}.
The SUS features ten 5-point Likert scale statements that are filled out post-prototype exposure.
The reported answers are calculated into an interpretable score between 0 (negative) and 100 (positive).
To further assist the numerical result interpretation, we also consider the adjective ratings as proposed by \citet{bangor2009dwi}.  
Furthermore, we also intended to make an assessment of the user's engagement as part of their overall experience when operating the implemented 3D UI.
For that purpose, we utilized the \textit{User Engagement Scale - Short Form} (UES-SF) questionnaire \citep{obrien2018apa}, and also completed post-prototype exposure.
The UES-SF features twelve 5-point Likert scale statements across four factors (three per factor): \textit{Focused Attention}, \textit{Perceived Usability}, \textit{Aesthetic Appeal}, and \textit{Reward}. 
Received answers may be scored in regard to the respective factors and as a combined user engagement score. 
Finally, we also integrated a logging system in the prototype, allowing for the recording of all user interactions with a respective timestamp.

During the task completion, the researcher made observations and took notes about to the user's interaction.
After task and questionnaire completion, a brief semi-structured interview with each participant was conducted, comprised of the following steps:

\begin{itemize}
    \setlength\itemsep{0em}
    \item Introductory preface: \textit{3D gestural input, or maybe more commonly referred to as ``hand interaction'', allows you to interact in a virtual environment, for instance by directly grabbing and manipulating virtual objects, or by making hand postures and gestures that are associated with certain features.}
    \item Q1: \textit{How do you feel about hand interaction that allows such an interaction in virtual reality?}
    \item Q2: \textit{In regard to the experienced prototype, what is your impression of how the hand interaction was implemented there?}
    \item Additional open remarks and comments, potentially based on the observations made during the task completion.
\end{itemize}

\subsection{Study Procedure}

The same procedure of five stages was followed in each study session:\footnote{The overall study duration was aimed to take 50--60 minutes, whereof the participants spend around 25 minutes wearing the HMD. The minute statements as listed were approximates, and naturally varied between participants.}

\begin{enumerate}
    \item Introduction (10~min);
    \item Warm-up (5~min in VR);
    \item Task (20~min in VR);
    \item Questionnaires (5~min);
    \item Interview (10~min).
\end{enumerate}

The participant first filled out an informed user consent, after which some demographic information was inquired (professional background and prior VR experiences).
Afterwards, the researcher introduced the overall context, scenario, and prototype including all its interactive features (via pre-recorded video). 
Each participant was then given some warm-up time with the prototype, i.e., they could get comfortable wearing the HMD, and familiarize themselves with the composition of the VE and the 3D gestural input.
Once they felt ready, the researcher initiated the task stage as described in Section~\ref{sec:task}.
To avoid a potential insights transfer from the warm-up to the task stage, different datasets were used.
Participants completed tasks one by one until all were completed (see~\autoref{tab:taskseries}).
The researcher observed the participant in the physical real-world space as well as in the VE from their HMD point of view as mirrored to a screen on the researcher's workstation, and took notes.
The researcher read aloud the individual tasks, and noted the participant's answers, likewise stated aloud. 
Once all tasks were completed, the participant was asked to complete in order the SUS and UES-SF questionnaires.
Finally, a semi-structured interview was conducted, after which they were thanked and sent off.

\subsection{Ethical Considerations}

We followed general ethical guidelines for the work with human participants within the scope of human-computer interaction research \citep{nent2016gfr,swedishresearchcouncil2017grp}.
The presented empirical evaluation was conducted between April and June~2021 during the, at the time ongoing, global COVID-19 pandemic, requiring the implementation of additional practical precautions.
All national, regional, and local health/safety recommendations according to the respective authorities were closely monitored and followed.
Study sessions were only conducted when the researcher and participant were symptom-free.
The researcher and participant kept recommended physical distance at all times during the study session.
The researcher was wearing a face mask at all times.
Each participant was provided with free access to face masks and hand disinfection gel.
All technical equipment was carefully sanitized between study sessions.

\section{Results}\label{sec:results}

\subsection{Participants}
\begin{figure}
        \begin{center}
            \includegraphics[width=.9\columnwidth]{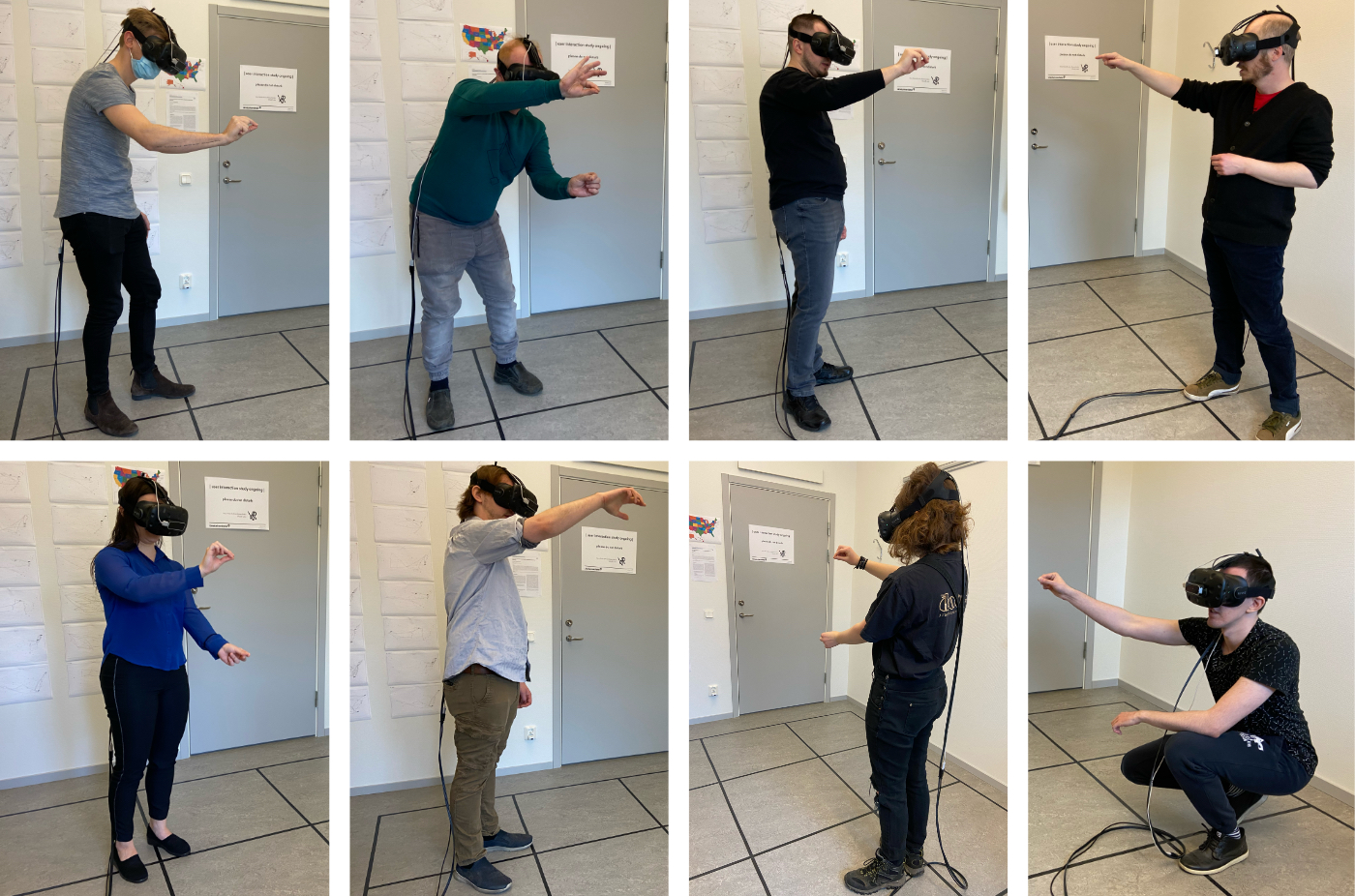}
        \end{center}
    \caption{Immersed participants during their task completion, wearing an HMD and interacting in the VE with the designed VR prototype as described throughout Section~\ref{sec:vrprototypeand3duidesign}.}\label{fig:evalparticipants}
\end{figure}

We recruited a total of $n = 12$ participants, reporting a variety of backgrounds: 5~Computer and Information Science, 5~Linguistics and Language Studies, and 2~Forestry and Wood Technology.
Eight participants stated \textit{a little}, three \textit{average}, and one \textit{a lot} prior experiences with VR.
None of them reported any visual perception issues when asked during the warm-up phase, e.g., in regard to their ability to differentiate the five Data Variable Axes of a 3D Radar Chart.\footnote{Applied color coding throughout the VE adopted from \href{https://colorbrewer2.org/}{colorbrewer2.org}.}
\autoref{fig:evalparticipants} presents some participant impressions.

\subsection{Task}

All participants were able to successfully complete the tasks (see Section~\ref{sec:task}) and provide correct answers as pre-determined, or otherwise contextually appropriate based on their own selection choices (tasks T06, T07, T08, T10, T14, T15, T23, T28, and T30).
According to the log file analysis, the task duration times averaged with $M = 13.95~min$ ($SD = 3.15~min$; tasks were presented in a swift manner without noticeable breaks; participants were instructed to complete them at their own pace).
When the participants were asked to select a time event that they deemed as ``\textit{interesting}'' and to briefly describe why (T14 and T28), they made their own observations, generally ending up selecting time events that featured either comparatively high or low data variable values.
These time events were visually noticeable, allowing them to make comparisons and to begin speculating for potential reasons.
Such participant descriptions included:

\begin{itemize}
    \setlength\itemsep{0em}

    \item ``\textit{Berries are very low, while Bananas and Oranges are high. This could indicate a different season of the year, thus the values across the different dimensions represent a change of season.}'' (T14, P1, day 75)
    
    \item ``\textit{Oranges and Bananas appear to be very high, while Grapes and Apples are very low. It seems like there is a relationship between those, maybe a seasonal event.}'' (T14, P7, day 132)
   
    \item ``\textit{Berries are very low, and then increasing afterwards. This is interesting, what is happening here.}'' (T14, P9, day 72)
    
    \item ``\textit{Oranges and Bananas are very high, while the others are very low. This looks like opposite trends.}'' (T14, P10, day 126)

    \item ``\textit{Oranges appear to be very high compared to the time series before and after the selected time event, maybe this could be because of a seasonal effect.}'' (T28, P2, day 58)
    
    \item ``\textit{The values \dots~seem to be at their dimension's average at the same time. It's a perfect overlap.}'' (T28, P4, day 86)
    
    \item ``\textit{Peak in the Grapes dimension, and it seems that Grapes are generally rather low overall compared to all other dimensions, therefore this is interesting.}'' (T28, P6, day 133)

    \item ``\textit{Grapes are high and we are in Italy, so this should be great for the wine season.}'' (T28, P12, day 145)
\end{itemize}

\subsection{Usability and User Engagement}

\begin{figure}
    \begin{center}
    \includegraphics[width=.75\columnwidth]{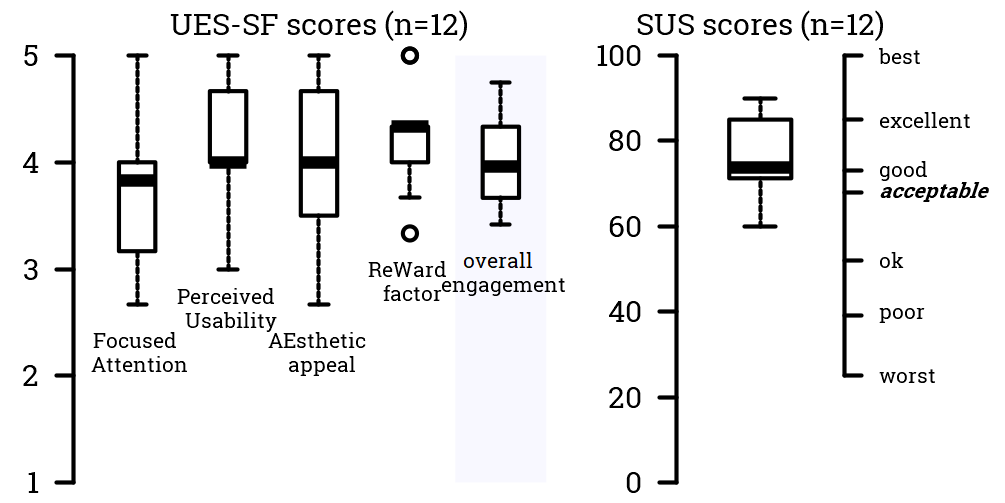}
    \end{center}
    \caption{\textbf{Left:}~The results of the UES-SF, presented according to the different engagement dimensions and the overall user engagement. The medians for each individual factor score (incl. overall engagement) are above average.
    \textbf{Right:}~The results of the SUS, presented including the original numerical scale and the supplemental adjective ratings according to \citet{bangor2009dwi}: worst~(25), poor~(39), ok~(52), good~(73), excellent~(85), best~(100). The mean value ($M=76.25$, $SD=9.62$) is well above the acceptable~(68) threshold.
    }\label{fig:SUS_UES-SF}
\end{figure}

\autoref{fig:SUS_UES-SF} presents the UES-SF and SUS scores.

\subsection{Observations}

For the most part, all participants appeared to understand the concept and learn the operation of the implemented features rather quickly, allowing them to interact in the VE seemingly naturally and in an enjoyable manner.
Nevertheless, some interesting observations were made throughout all the study sessions, summarized as follows.

\subsubsection{Usability Issues}\label{sec:results_usabilityissues}

Most noticeably, the time navigation (\textit{Time Event Selection} by grabbing, dragging, and releasing a 3D Radar Chart's Time Slice) appeared comparatively sensitive during the interaction's conclusion.
The participants had seemingly no problems initiating and continuing the grabbing mechanic, navigating back and forth in time while simultaneously interpreting the data and reading the updated labels in the juxtaposed Information Window.
However, when asked to select a specific time event (T04, T22, and T29), at times the Time Slice would snap to an adjacent time event during the hand-based grasping's release.
By opening up one's hand, the hand tracking would first interpret a Time Slice movement before concluding the grasping gesture and discontinuing the time navigation.
In these cases, participants had to attempt this interaction more than once until the Time Slice remained in the desired position.
Such reoccurring observations were made during nine study sessions.

The \textit{Zoom (in/out)} gestural command seemed to require the comparatively longest learning phase. 
Depending on a participant's hand placement, the tracking sensor would sometimes discontinue detecting the lower hand, as it appeared to be (partially) occluded by the hand above.
Once the participants appeared to have gotten a more cautious understanding and feeling of the hand tracking, they were able to perform these gestural commands seemingly fluently.
One participant was observed repeatedly attempting the gestural commands in their reverse concept, i.e., moving hands together to \textit{Zoom in}, and moving hands apart to \textit{Zoom out}.

Some instances of unintentional commands were observed, i.e., a participant triggered a feature through the 3D gestural input without explicit intent.
Most noticeably, this occurred during intended \textit{Mode Toggle} interaction, resulting in unintended \textit{Travel}.
In these cases, rather than touching the 3D Radar Chart's Activation and Interaction Toggle with a hand and all fingers extended, the participant would attempt to touch it with only the index finger extended, similar to a ``poking''-like posture.
This however was in conflict with the same hand posture configuration as part of the \textit{Travel} feature's gestural command, thus resulting in an unintentional movement.

\subsubsection{General Operation and Interaction}

To make data observations, the participants appeared to use a balanced mixture of actively moving around a 3D Radar Chart and in-place \textit{Rotation} using its Rotation Handle. 
Even though not explicitly asked, some participants made on their own accord noticeable use of various implemented features to assist them with their task-solving process, e.g., sorting the data variables before selecting a time range (T08 and T10), or filtering out proclaimed ``\textit{uninteresting}'' data variables (T14 and T28). 
The participants were asked to sort the data variables in ascending (T15) and descending (T30) order.
However, at no point were they told what these orders mean within the presented context.
We were curious to observe how the participants themselves interpreted these tasks.
The majority associated \textit{ascending} with a \textit{clockwise} and \textit{descending} with a \textit{counter-clockwise} radial arrangement of the data variables with respect to their visualization in a 3D Radar Chart's Information Panel. 
A few participants appeared rather self-critical with their perceived performance operating the 3D UI, but became seemingly more confident over time as they got ``\textit{a better feeling}'' for the hand tracking.
Sometimes, participants attempted to perform gestural commands rather quickly, while their hands were not yet in the tracking sensor's field of view.
Although their gestural input operation was correct in concept, the tracking sensor appeared too slow in its initial hand detection, thus preventing them from the practical execution of the respective interaction.
This was frequently observed for those features classified as \textit{gestural commands}, but not so much for the \textit{hand-based grasping} ones.

\subsection{Interview}

\subsubsection{General Hand Interaction in VR}

When asked how the participants feel about using their hands as means of interaction in VR (Q1), they expressed a rather positive attitude towards it.
They thought that hand interaction has the potential to allow for very natural and intuitive interaction mechanisms.
Some of them mentioned their appreciation that no additional sensors needed to be attached to one's hands.
One participant expressed minor concerns about imprecise command recognition: When an interaction is not triggered, even though correct in concept, it might make the user feel insecure, as it is difficult to determine whether the detection problem was due to them or the system. 
Four participants explicitly expressed their appreciation for simply using their hands instead of physical controllers that can ``\textit{sometimes feel weird for the interaction, as one is grabbing a controller and the controller is grabbing a virtual object}'', therefore having some kind of middle layer impression -- which, according to them, is not the case with hand interaction.

\subsubsection{3D UI of the Prototype}

When asked how the participants perceived the hand interaction within the scope of the implemented prototype (Q2), they were generally positive about the provided features.
The majority stated that the 3D UI felt very natural and easy to operate once one had learned all possibilities.
They acknowledged their impression of learning the various features quickly, with one participant elaborating that it felt like ``\textit{riding a bike}'' at that stage.
Some of them noted that the 3D UI featured logical and coherent analogies for the different hand postures and gestures.
A few were genuinely surprised that seemingly many features relied on the utilization of both hands simultaneously, expecting more one-handed gestures.
Participants also addressed some of the encountered usability issues, most dominantly mentioning that the precise Time Slice placement appeared to be ``\textit{fairly tricky}'' at times (as described in Section~\ref{sec:results_usabilityissues}), making it feel as if the hand tracking was too sensitive in these instances.
Some also reflected on experiencing unintentional gestural commands.

\section{Discussion}\label{sec:discussion}

Generally, all study participants were able to interact organically and intuitively in the immersive VE using the implemented 3D gestural interface, having a smooth and responsive experience with the prototype.
In contrast to the gestural control results reported by \citet{streppel2018iiv}, the majority of our participants managed to learn the features of the 3D UI comparatively quickly, both conceptually and operationally, completing the different tasks they were presented with.
\citet{huang2017ags} reported similar subjective impressions towards learnability and intuitiveness based on the evaluation of their prototype.
When asked to do a certain action within the task series, our participants were able to quickly associate the correct interaction in VR, i.e., the visual object they had to manipulate or the hand posture/gesture they had to perform.
The median and mean scores of the measured usability (SUS) were above the \textit{good} threshold.
Given our focus on hand-based grasping and gestural command techniques, we are overall satisfied with the results considering the participants were asked to conduct a multitude of predefined tasks rather than just freely exploring the data at their own leisure.
The overall user engagement scores (UES-SF; between 3 and 5, median slightly below 4) are also encouraging, indicating positive engagement with the prototype by the participants.
This aligns with our observations as they would often use features such as \textit{Rotation}, \textit{Sort}, and \textit{Filter}, even when not explicitly asked for, seemingly naturally engaging with the prototype.
A closer examination of the individual engagement factor scores reveals indications that the participants paid close attention during the task completion, perceived the usability as good (in allignment with the SUS scores), found the prototype aesthetically appealing, and their experience rewarding (with medians around 4) -- all in anticipation of its general design objective. 

\subsection{Reflection: Hand-based Grasping Interaction}

A major aspect of the 3D gestural interface's design was concerned with the utilization of hand-based grasping for the interaction with \textit{visible virtual objects} in the VE, which was appreciated by the participants.
They were able to interact with the Axis Spheres of the Reconfigure and Filter Handle as an indirect widget to adjust the configuration of the 3D Radar Chart, similar to the node movement interaction as demonstrated in the prototypes by \citet{osawa2000ign} and \citet{huang2017ags}.
They could intuitively grab and drag the Time Slice in order to make respective \textit{Time Event Selections}.
While this interaction was valued, some shortcomings were identified when the participants had to place the Time Slice at a specific time event.
The tracking and implementation felt ``\textit{too sensitive}'' as the Time Slice would sometimes ``snap'' into one of the adjacent time events when attempting to release the grab, occasionally resulting in light frustration and requiring some additional interaction to recover from this error -- a cost that should not be ignored at a larger scale \citep{buschel2018ifi}.
The Time Slice movement is dependent on the detected grab-position of the hand, i.e., the position where fingers and thumb meet.
In the process of releasing the grab, this position is likely to be updated slightly before the grab is detected as discontinued, thus no longer updating the time event selection.
Based on the current implementation, this issue is proportionally dependent on the length of the 3D Radar Chart and the amount of included time events, i.e., the resolution of time events.
As a reference, a 3D Radar Chart was scaled to correspond to a total length of 100~cm in the VE, with a total of 150~time events encoded, resulting in an effective gap distance between two time events of 0.67~cm.
A lower amount of included time events over the same length would result in a larger gap between individual time events (as for instance in the case when \textit{Zoomed in}), which would prevent the Time Slice from snapping to an adjacent time event accordingly.
Vice versa, including even more events in the time series, would further increase the perceived sensitivity.
While we expect 3D gestural input technologies to become more precise, we also envision some solutions based on the overall 3D UI design and implementation.
For instance, rather than exclusively relying on the finger and thumb positions for grab detection, one could implement an additional dependency based on the hand's back or palm position.
In the presented case of grabbing and vertically dragging the Time Slice, the hand's back and palm position are likely to remain relatively static in space during the release of the hand-based grasping compared to finger and thumb movements.
A threshold could be implemented to prevent Time Slice movement in such instances, enabling the system to ``interpret'' the user's intention to discontinue their interaction.
Alternatively, another approach to solving this challenge could be based on an asymmetric bimanual interaction, similar to as presented in the prototype by \citet{betella2014uln}.
For instance, while grasping the Time Slice with one hand, a gestural command made with the other could ``lock'' the current Time Slice position in place, allowing to safely disengage from the interaction without unintentionally moving forward or backward in time.

\subsection{Reflection: Gestural Commands}

In addition to interacting with visual objects, we also implemented a set of \textit{invisible gestural commands} in the 3D UI.
Gestural commands such as for \textit{Travel} and \textit{Time Range Selection} were positively received.
The participants appreciated the responsiveness of the \textit{Time Range Selection}, allowing them to live highlight the time ranges they were interested in.
The continuous semi-transparent uncolored visualization of the time series data outside these ranges provided them with a further preview of the data, which was particularly important for them when making the cutoff and deciding whether or not to include additional time events in their selection.
The two-handed gestural commands worked generally well.
However, based on our observations and the received feedback, some improvements can be made in regard to the \textit{Zoom (in/out)} feature.
In the initial hand posture of holding both hands vertically slightly apart with their palms facing each other, the tracking sensor sometimes did not recognize the lower hand as it was occluded through the one above.
Thus, even though the participants were holding their hands in the correct configuration, they needed to move them around slightly before the sensor tracked and translated them appropriately in the VE.
Similar feedback was stated by the participants in the evaluation as reported by \citet{huang2017ags}, expressing a desire for more robust gesture recognition in such instances.
Moving both hands together and then apart, or vice versa, for zoom operations was also reportedly preferred as an interaction design approach by the participants in the study by \citet{austin2020esi}.
Both our findings as well as the previously described ones by \citet{huang2017ags} highlight thus the importance of a reliable implementation of such bimanual interactions in the future to further satisfy anticipated user preferences.

\subsection{Reflection: Unintentional Commands}

Cases of unintentional gestural commands \citep{pavlovic1997vio} occurred most noticeably when a user wanted to display details-on-demand by touching a 3D Radar Chart's \textit{Mode Toggle} widget, but instead triggered a \textit{Travel} interaction, as their hand posture was detected as pointing forward.
While participants were able to travel back and recover from such an error comparatively quickly, it also caused them a mixture of light surprise, frustration, and uncertainty towards the \textit{Mode Toggle} interaction.
This is a great example of such an unintentional command, demonstrating that different users may attempt the same interaction differently in regard to their hand posture.
We envision that such an issue can be fixed based on the current implementation in various ways, e.g., through the implementation of a distance threshold between the virtual hand model and the \textit{Mode Toggle} widget, i.e., preventing \textit{Travel} if a user's hand is detected in close proximity to the widget.
Thus, the 3D UI may infer in-situ that the user intends to engage with a 3D Radar Chart rather than attempting to \textit{Travel}.
We reflect on this practical example by highlighting again the discussion by \citet{nehaniv2005ama} about the importance of a computing system's ability to infer the user's intent with their interactions.
On the other hand, no unintentional \textit{Reset} operations were observed, even though the participants were able to perform the command swiftly.
Similar to the considerations by \citet{fittkau2015esc}, we intentionally designed this command to prevent unintentional performance, as resetting a 3D~Radar Chart's configuration is a comparatively drastic operation.

\section{Conclusion and Future Work}\label{sec:conclusion}

The designed and implemented 3D gestural interface allowed our study participants to interact with spatio-temporal data in an immersive VE to complete a series of typical analytical tasks.
We described the 3D UI design and its features within the context of IA, informed by relevant foundational work, such as data analysis tasks (see Section~\ref{sec:vrprototype_adopteddataanalysistaskterminology}, 3D interaction techniques \citep[Chapters~7--9]{laviola20173ui}, and aspects of hand posture comfort \citep{rempel2014tdo}.
The results of our empirical evaluation with $n = 12$ participants point towards good usability and an overall engaging experience, where the participants were excited to intuitively use their hands to operate the VR prototype using hand-based grasping and gestural commands to interact with the abstract data visualizations as 3D~Radar Charts \citep{reski2020eot}.
We discussed the results and were able to reflect on the 3D UI design, identifying aspects for improvement related to hand tracking detection and precision as well as a VR system's ability to infer user intent to avoid unintentional gestural commands.
Even though tracking sensors are likely to improve, we envision that most if not all of these aspects can be addressed through careful design and implementation on the software side.

In addition to the study presented here, we also utilized the presented 3D UI design\footnote{All features were available to the users besides the \emph{Zoom (in/out)}, which was excluded due to the design of the collaborative study tasks.} in a hybrid asymmetric collaborative study setup, involving both an immersed and a non-immersed user \citep{reski2022aee}. 
This follow-up study differed in various aspects compared to the one presented in this article; instead of interacting with the prototype in a walkthrough-like manner, the immersed users interacted on their own accord, in sessions that lasted approximately twice as long, to explore the data and solve confirmative data analysis tasks.
Although not directly comparable to this study, among other obtained results, the usability and user engagement were rated similarly positive based on the administered SUS and UES-SF.

We are generally satisfied with the outcome of the presented work and have some ideas for future iterations.
For instance, we are motivated to improve the prototype based on the discussed aspects and investigate its application within the scope of longitudinal studies. 
Under the assumption of being immersed in the VE for a longer duration, it then also makes sense to investigate explicitly aspects of the 3D gestural interface's comfort and physical fatigue.
Even though the participants in our study were able to quickly learn the operation of the 3D UI, it would also be intriguing to investigate more specifically learnability aspects -- a topic that is often disregarded and underexplored \citep{rempel2014tdo}.

\section*{Acknowledgments}
\noindent
The authors wish to thank all the participants of the user interaction study.
This work was partially supported by the ELLIIT environment for strategic research in Sweden.

\section*{Author Contributions}
\noindent
All authors (NR, AA, and AK) devised the research scope.
NR and AA designed the empirical evaluation.
NR reviewed the literature, developed all technical parts of the immersive VR environment, recruited study participants, and conducted the empirical evaluation (user interaction study) as well as the data collection.
AA and NR conducted the data analysis.
NR and AA wrote the manuscript.
All authors discussed and reviewed the manuscript.

\bibliography{references02.bib}

\end{document}